\documentclass[aps,prl,twocolumn,superscriptaddress,showpacs]{revtex4}
\usepackage[intlimits]{amsmath}
\usepackage{amsfonts}
\usepackage{graphics}
\usepackage{subfigure}
\usepackage[usenames]{color}
\usepackage{epstopdf,epsfig}

\usepackage{indentfirst}

\usepackage{graphicx}
\usepackage{amsmath}
\usepackage{amsfonts}
\usepackage{ifpdf}

\ifpdf
\usepackage{epstopdf}
\usepackage{hyperref}
\else
\usepackage[hypertex,ps2pdf,dvips,dvipdf,colorlinks,urlcolor=blue,citecolor=blue]{hyperref}
\fi

\newcommand\be            {\begin{equation}}
\newcommand\bea           {\begin{equation}\begin{array}l\displaystyle}
\newcommand\ee            {\end{equation}}
\newcommand\bes           {\begin{subequations}}
\newcommand\esu           {\end{subequations}}
\newcommand\erf[1]        {\eqref{#1}}
\newcommand\labl[1]       {\label{#1}\ee}

\newcommand{\ud}{\mathrm d}

\newcommand\fii           {\varphi}
\newcommand\mc            {\mathcal}

\newcommand\p            {\partial}
\newcommand\psid         {\psi^{\dagger}}

\newcommand\vev[1]{{\langle#1\rangle}}

\newcommand\no[1]{{\,:\!#1\!:\,}}
\def\3pt#1#2#3{{\langle{#1}\vert{#2}\vert{#3}\rangle}}

\newcommand\doi[2]        {\href{http://dx.doi.org/#1}{#2}}
\newcommand\httpurl[2]    {\href{#1}{#2}}

\begin{document}

\title{Expectation Values in the Lieb--Liniger Bose Gas}

\author{M. Kormos}
\affiliation{SISSA and INFN, Sezione di Trieste, via Beirut 2/4, I-34151, 
Trieste, Italy}

\author{G. Mussardo}
\affiliation{SISSA and INFN, Sezione di Trieste, via Beirut 2/4, I-34151, 
Trieste, Italy}
\affiliation{International Centre for Theoretical Physics (ICTP), 
I-34151, Trieste, Italy}

\author{A. Trombettoni}
\affiliation{SISSA and INFN, Sezione di Trieste, via Beirut 2/4, I-34151, 
Trieste, Italy}


\begin{abstract}
\noindent 
Taking advantage of an exact mapping between a relativistic integrable
model and the Lieb--Liniger model we present a novel method to compute expectation values in the Lieb--Liniger Bose gas both at zero and finite temperature.
These quantities, relevant in the physics of one-dimensional ultracold Bose gases, are expressed by a series that has a remarkable behavior of convergence. 
Among other results, we show the computation of the three-body 
expectation value at finite temperature, a quantity that rules the 
recombination rate of the Bose gas. 
\end{abstract}


\maketitle

Correlation functions are key quantities in quantum interacting systems: 
not only they fully encode the dynamics but 
also are directly related to various susceptibilities and response functions. 
For these reasons, 
there has always been an intense search to find 
the most efficient ways to compute them. The task is notoriously difficult, even if the system is integrable. A typical but significant example is provided 
by the Lieb--Liniger (LL) model \cite{LL} that 
describes the low-temperature properties 
of one-dimensional interacting Bose gases: 
although it can be solved through Bethe Ansatz equations, the explicit computation of its correlation functions is a long-standing problem \cite{korepin, weiss}. The interest for the computation of correlation functions in the LL model 
is obviously not only theoretical. 
In a series of recent experimental achievements strongly interacting 
ultracold bosons have been confined within wave-guides by nearly one-dimensional potentials that tightly trap the particle motion in the two transverse directions while leaving it free in the third axial direction \cite{kinoshita04,paredes04,vandruten08}:  
the coupling of these bosons with the external world can be made so weak that their behaviour is very well described by the LL
model. Through interference (or eventually {\it in situ}) experiments, 
several quantities can be detected both at zero and finite temperature: the time duration of experiments depends on the three-body recombination 
rate, which is proportional to local three-body expectation values \cite{kagan}. Many important general questions of quantum many-body physics
can be studied in such a highly controllable set-up:  dynamical properties concerning the absence (or not) of thermalization \cite{kinoshita04,olshanii2}, for instance, or the behaviour of integrable quantum systems 
when small non-integrable perturbations (e.g., three-body interactions and/or a weak external trapping potential) are switched on \cite{DMS}.

\vspace{-1mm}
Over the years several theoretical quantities of the LL model have been computed by means of different techniques   
\cite{yang,haldane,slavnov,giorgini,gangardt,kheruntsyan,cazalilla,castin,deuar,cheianov}.  
In this paper we present a compact and general way to determine the expectation values of its local operators. The method takes advantage of an 
exact mapping between a relativistic integrable massive model -- the sinh--Gordon (ShG) -- and the LL model: in a proper non-relativistic limit of the ShG model, both its $S$-matrix and Lagrangian coincide with those of the LL model. 
Since the $S$-matrix of an integrable relativistic model fixes the exact matrix elements of all operators of the theory (and for the ShG model these matrix elements are all known \cite{FMS,mussardo}), the correspondence between the two models opens the way to computing the corresponding quantities of the LL model in a very direct way. 
 As shown below, this method provides a remarkable simplification of the problem. Its implementation actually requires to take into account an additional aspect of the problem: while in the ShG model the
correlation functions refer to the vacuum (i.e. the state without any particles), in the LL model they relate instead to its ground state at a finite density. This aspect, however, can be successfully overcome 
by the Thermodynamical Bethe Ansatz (TBA) formalism developed in \cite{leclair}, which has the additional convenience of being applicable equally well both at zero and finite temperature. In this way we are able to compute not only the zero temperature expectation values but also their finite temperature expressions. 

{\em The LL-ShG mapping}.
The LL Hamiltonian for $N$ interacting bosons of mass $m$ in one dimension is 
\be
\label{HAM_LL}
H\,=\,-\frac{\hbar^2}{2m}\sum_{i=1}^N\frac{\p^2}{\p x_i^2}+2\lambda\,
\sum_{i<j}\delta(x_i-x_j)\,.
\ee
The corresponding non-relativistic field theory description is the quantum non-linear Schr\"odinger model \cite{korepin}, which 
employs the complex field $\psi$ and the Lagrangian  
\be
{\cal L}= -\frac{\hbar^2}{2m}|\nabla\psi|^2
+ 
i\,\frac\hbar2\left(\psid\frac{\p\psi}{\p t} - \frac{\p\psid}{\p
  t}\psi\right) - \lambda\,|\psi|^4\,.
\label{LagrangianNLS}
\ee
The effective coupling constant of the LL model is given by the dimensionless parameter $\gamma=2m\lambda/\hbar^2n$, where $\lambda > 0$ is the coupling entering the Hamiltonian (\ref{HAM_LL}) while $n=N/L$ is the density of the gas ($L$ is the length of the system). Temperatures
are usually expressed in units of the temperature $T_D=\hbar^2 n^2/2mk_B$ of the quantum degeneracy, $\tau=T/T_D$. The two-body elastic S-matrix of the LL model is \cite{LL,yang}
\be
S_\text{LL}(p,\lambda)\,=\,\frac{p-i 2m \lambda/\hbar}{p+i 2m \lambda/\hbar}\,,
\label{eq:SLL}
\ee
where $p$ is the momentum difference of the two particles. 

Consider now the ShG model in (1+1) dimensions, i.e.\ the integrable and relativistic  invariant field theory defined by the Lagrangian 
\be
\mc{L}_\text{ShG}= \frac12\left[\left(\frac{\p\phi}{c\,\p
  t}\right)^2-\left(\nabla\phi\right)^2\right] -
\frac{\mu^2}{g^2}
\cosh(g\phi)\,,
\label{LagrangianShG}
\ee
where $\phi=\phi(x,t)$ is a real scalar field, $\mu$ is a mass
scale and $c$ is the speed of light. The parameter $\mu$ is related to the physical (renormalized) mass $m$ by $\mu^2=\pi \alpha m^2c^2 / \hbar^2 \sin(\pi\alpha)$, where $\alpha=\hbar c\,g^2 / (8\pi+\hbar c\,g^2)$ \cite{FMS}. The energy $E$ and the momentum $P$ of a particle can be written as $E=m c^2 \cosh\theta$, $P=m c \sinh\theta$, where $\theta$ is the rapidity.  
Since the ShG dynamics is ruled by an infinite number of conservation laws, all its scattering processes are purely elastic and can be factorized in terms of the two-body S-matrices \cite{FMS}
\be
S_\text{ShG}(\theta,\alpha)\,=\,\frac{\sinh\theta-i\,\sin(\alpha\pi)}{\sinh\theta+
i\,\sin(\alpha\pi)}\,,
\label{eq:SH-G}
\ee
where $\theta$ 
is the rapidity difference of the two particles. It is now easy to see that taking simultaneously the non-relativistic and weak-coupling limits of the ShG model such that  
\be
g\to0,\;c\to\infty,\quad g\,c=4\sqrt{\lambda}/\hbar=\text{fixed}\,, 
\label{eq:limit}
\ee
its $S$-matrix (\ref{eq:SH-G}) becomes identical to the $S$-matrix (\ref{eq:SLL}) of the LL model. Notice that the coupling $\lambda$ does not need to be small, i.e.\ with this mapping we can study the LL model at arbitrarily large values of the dimensionless coupling $\gamma$.

The mapping between the two models goes beyond the identity of their $S$-matrix: it actually extends both to their Lagrangians and TBA equations. Details will be given elsewhere 
, but it is simple to follow the main steps of the procedure. 
According to \cite{beg}, the non-relativistic limit of a field theory consists of expressing the real scalar field in the form 
\[
\phi(x,t)=\sqrt{\frac{\hbar^2}{2m}}\left(\psi(x,t)\,
  e^{-i\frac{mc^2}\hbar\,t}+\psid(x,t) e^{+i\frac{mc^2}\hbar\,t}\right)\,,
\]
and, when the limit $c\to\infty$ of the Lagrangian is taken, of omitting all its oscillating terms 
. The commutation relation
$
[\phi(x,t),\Pi(x',t)] = i\hbar\,\delta(x-x')
$
implies for the non-relativistic operators
$
[\psi(x,t),\psid(x',t)]=\delta(x-x')\,.
$
Furthermore, when the limit $g \to 0$ of 
eqn \erf{eq:limit} is considered, the $\psid\psi$ terms  coming from
the potential and kinetic parts cancel each other, while all higher
terms of the series expansion of the potential, but the quartic one, vanish. Hence, the ShG Lagrangian (\ref{LagrangianShG}) reduces
to the non-linear Schr\"odinger Lagrangian (\ref{LagrangianNLS}). Notice that the mapping based on the limit \erf{eq:limit} applies to any operator of the theory. 

In the same way one can also show that the TBA equations of the ShG model (given for instance in 
\cite{klass}) reduces to the ones of the LL model, written down in \cite{yang}. In the LL model at a finite $T$ 
the TBA equation 
for the pseudo-energy $\epsilon(T,\mu)$ consists of the non-linear integral equation 
\be
\epsilon(T,\mu)\,=\, \frac{p^2/2m -\mu}{k_B T} - \fii \circ 
\log\left(1+e^{-\epsilon}\right)\,,
\label{eq:TBA}
\ee
where $\mu$ is the chemical potential associated to the finite density $n$ of the gas, $\fii(p) = - i\frac\p{\p p}\log S_{LL}(p)$ 
is the derivative of the phase shift and 
$\varphi \circ f \equiv \int_{-\infty}^\infty\frac{\mathrm{d} p'}
{2\pi}\,\varphi (p-p') f(p')$
. The solution of this integral equation leads to the free energy and to all other thermodynamical data of the model.   

{\em Expectation values}. At equilibrium the expectation value of an operator $\mc O = \mc O(x)$ at temperature $T$ and at finite density is given by 
\be
\vev{\mc O} = \frac{\mathrm{Tr}\left(e^{-(H-\mu N)/(k_\text{B}T)}\mc{O}\right)}
{\mathrm{Tr}\left(e^{-(H-\mu N)/(k_\text{B}T)}\right)}\,. 
\labl{eq:vev}
In a relativistic integrable model the above quantity can be neatly expressed as \cite{leclair} 
\be
\vev{\mc O} =\sum_{n=0}^\infty\frac1{n!}
\int_{-\infty}^\infty 
\left(\prod_{i=1}^n\frac{\ud\theta_i}{2\pi} f(\theta_i)
\right) 
\3pt{\overleftarrow{\theta}}{\mc O(0)}{\overrightarrow{\theta}}_\text{conn}\,,
\labl{eq:muss}
where $f(\theta_i) = 1/(1+e^{\epsilon(\theta_i)})$ and $\overrightarrow{\theta} \equiv 
\theta_1,\dots,\theta_n$ ($\overleftarrow{\theta} \equiv \theta_n,\dots,\theta_1$)
denote the asymptotic states entering the traces in (\ref{eq:vev}). This formula employs both the pseudo-energy $\epsilon(\theta)$ and the connected diagonal form factor of the operator ${\mc O}$, defined as 
$
\langle \overleftarrow{\theta} \vert 
{\mathcal O} \vert \overrightarrow{\theta} \rangle_\text{conn}=
FP (\lim_{\eta_i\to0} 
\langle 0 \vert {\mathcal O} \vert  \overrightarrow{\theta}, 
\overleftarrow{\theta} - i\pi+i \overleftarrow{\eta} \rangle ) 
$
where $\overleftarrow{\eta} \equiv \eta_n,\dots,\eta_1$ and $FP$ in front of the expression means taking its finite part, i.e. omitting all the terms of the form $\eta_i/\eta_j$ and
$1/\eta_i^p$ where $p$ is a positive integer. 

In order to compute the expectation values of the LL model by applying eqn (\ref{eq:muss}) we need: (a) to solve the integral equation (\ref{eq:TBA}) for $\epsilon(\theta)$; (b) to identify the relevant form factors of the ShG model; (c) to take the non-relativistic limit of both the form factors and eqn (\ref{eq:muss}). Given for granted the straightforward (numerical) solution of eqn (\ref{eq:TBA}), let us focus our attention on the last two points. The generic $m$-particle form factor of a local operator ${\mc O}$ in the ShG model can be written as \cite{FMS,mussardo}
\be
F_m^{\mc O}(\theta_1,\dots,\theta_m) = \,Q_m^{\mc O} (x_1,\dots,x_m)\,\prod_{i<j}\frac{F_\text{min}(\theta_{ij})}{x_i+x_j}\,,
\ee
where $x_i=e^{\theta_i}$ and $Q_m^{\mc O}$ are the  symmetric polynomials in the $x$'s that fully characterize the operator ${\mc O}$. The explicit expression of $F_\text{min}(\theta)$ is given in \cite{FMS} but the only thing needed here is its functional equation 
\[
F_{\rm min}(i\pi+\theta) F_{\rm min}(\theta)=
\frac{\sinh\theta}{\sinh\theta+ i \sin(\pi \alpha)}\,.
\]
We are interested in the symmetric polynomials $Q_m^{(q)}$ of the
exponentials ${\mc O}_q = e^{q g \phi}$ since, using their Taylor expansion, we can extract the form factors of all normal ordered operators $\no{\phi^k}$. Their expression is \cite{mussardo} 
\be
Q_m^{(q)}= [q] \left(\frac{4\sin(\pi\alpha)}{N}\right)^{\frac{m}2}\det
M_m(q)\,,
\labl{eq:FFexp}
where $M_m(q)$ is an $(m-1)\times(m-1)$ matrix with elements $ \left[M_m(q)\right]_{i,j}=\sigma^{(m)}_{2i-j}[i-j+q]
$. Above, $[x]\equiv
\sin(x\pi\alpha)/\sin(\pi\alpha)$ while $\sigma^{(m)}_a$ ($a=0,1,\dots, m$) are the elementary symmetric polynomials in $m$ variables. 

\begin{figure}[t]
\centerline{
\scalebox{0.27}{\includegraphics{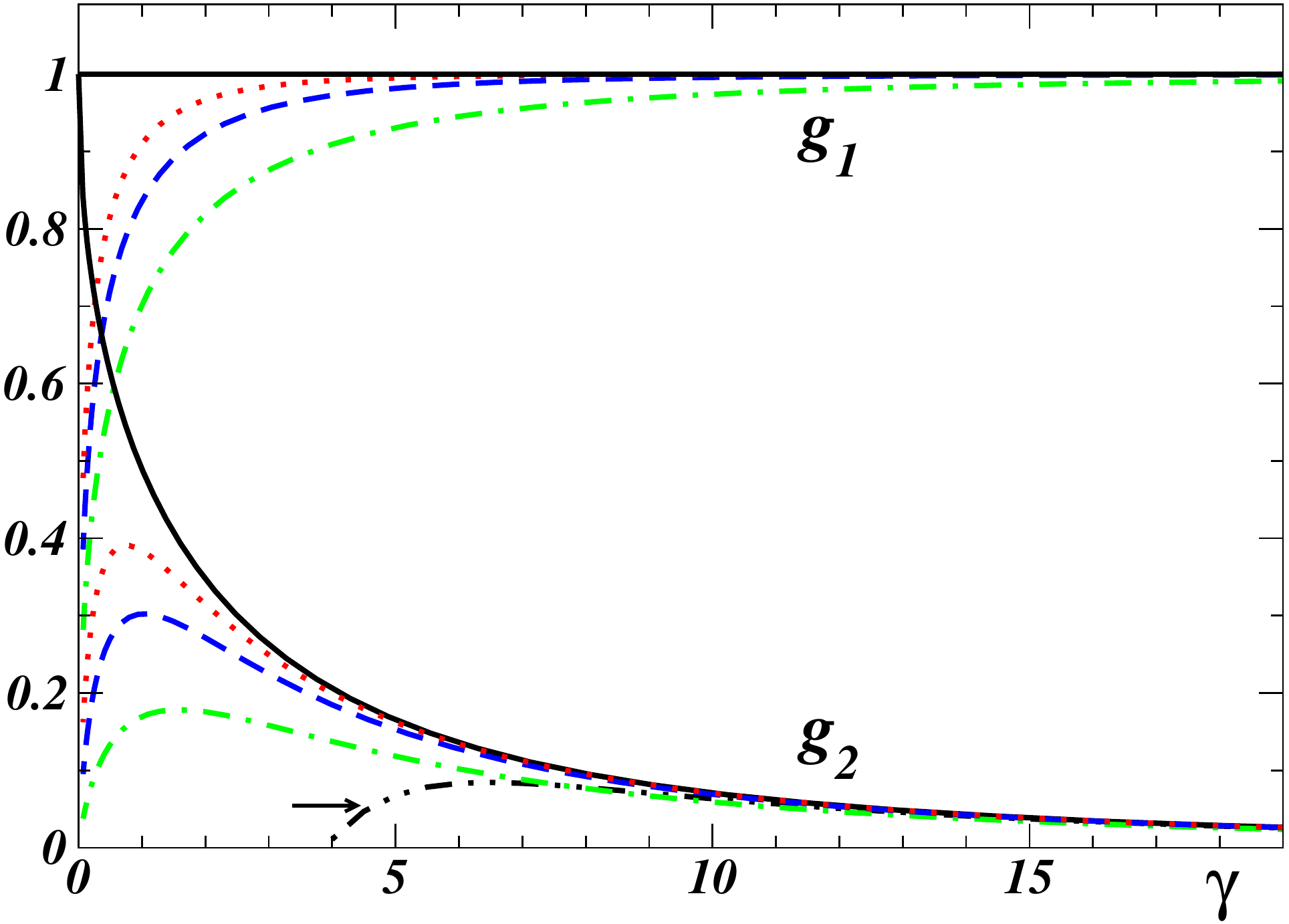}}
}
\caption{$g_1$ and $g_2$ at $T=0$ using form factors up to 
$n=\textcolor{green}{4}$, $\textcolor{blue}{6}$ and $\textcolor{red}{8}$ particles, 
respectively with green dot-dashed,  
blue dashed and red dotted lines. The exact values are given by the 
solid lines whereas 
the dot-dashed line below, indicated by the arrow,  corresponds to the strong coupling 
expansion (\ref{strong-coupling}).}
\label{fig:fig1}
\end{figure}

On the basis of the results given above, we are now in the position to compute the local $k$-particle correlation 
function $g_k$ of the LL model defined by 
\be
\vev{\psid\,^k\psi^k}=n^k\,g_k(\gamma,\tau)\,,
\ee
where $k$ is an integer ($k=1,2,3,\dots$). The $g_k$'s are functions of the dimensionless LL coupling $\gamma$ and of the reduced temperature $\tau$. 
The relation between $g_k$ in the LL model and the corresponding quantity in the ShG model  
in the limit (\ref{eq:limit}) is given by  
\[\vev{\no{\phi^{2k}}}\to 
\left(\frac{\hbar^2}{2m}\right)^k\binom{2k}{k}\vev{\psid\,^k\psi^k}\,.
\]
Using eqn \erf{eq:muss} and the connected form factors of the corresponding operator we arrive at the expression
\begin{gather}
\vev{\psid\,^k\psi^k} = 
 \binom{2k}{k}^{-1}\!\!\left(\frac{\hbar^2}{2m}\right)^{-k} 
\sum_{n=1}^\infty {\cal F}_n\,,
\label{eq:formula}\\
{\cal F}_n = \frac{1}{n!} \int_{-\infty}^{\infty} 
\left(\prod_{i=1}^n \frac{d p_i}{2\pi} f(p_i)\right)
\tilde F^{\no{\,\phi^k\,}}_{2n,\text{conn}}(p_1,\ldots,p_n)\,,\nonumber
\end{gather}
where
\[
\tilde
F^{\no{\,\phi^k\,}}_{2n,\text{conn}}(\{p_i\})=\lim_{c\rightarrow \infty,
g\rightarrow 0}\,\left(\frac1{mc}\right)^nF^{\no{\,\phi^k\,}}_{2n,\text{conn}}(\{\theta_i=\frac{p_i}
{mc}\})
\]
are the double limit (\ref{eq:limit}) of the
connected form factors. As shown below, the series (\ref{eq:formula})
are nicely saturated by the first few terms for sufficiently large
values of $\gamma$ ($\gamma = 0$ is a singular point of the model
\cite{LL}, therefore one cannot expect a priori any fast convergence nearby). 
A first check of eqn (\ref {eq:formula}) is provided by the case $k=1$: using (\ref{eq:formula}) (with a chemical potential $\mu$ that ensures the finite density $n$) and summing up the series, one easily checks that $\vev{\psid\psi}=n$ and $g_1=1$, as it should be for translational invariance. As shown in Fig.~\ref{fig:fig1}, the exact value $g_1=1$ (solid line) is rapidly approached by just the first terms of (\ref{eq:formula}): the convergence of the series is always remarkably fast for all $\gamma \geq 1.5$, where the exact value is obtained 
within a $5\%$ accuracy just using its first four terms. 
\begin{figure}[t]
\scalebox{0.23}{\includegraphics{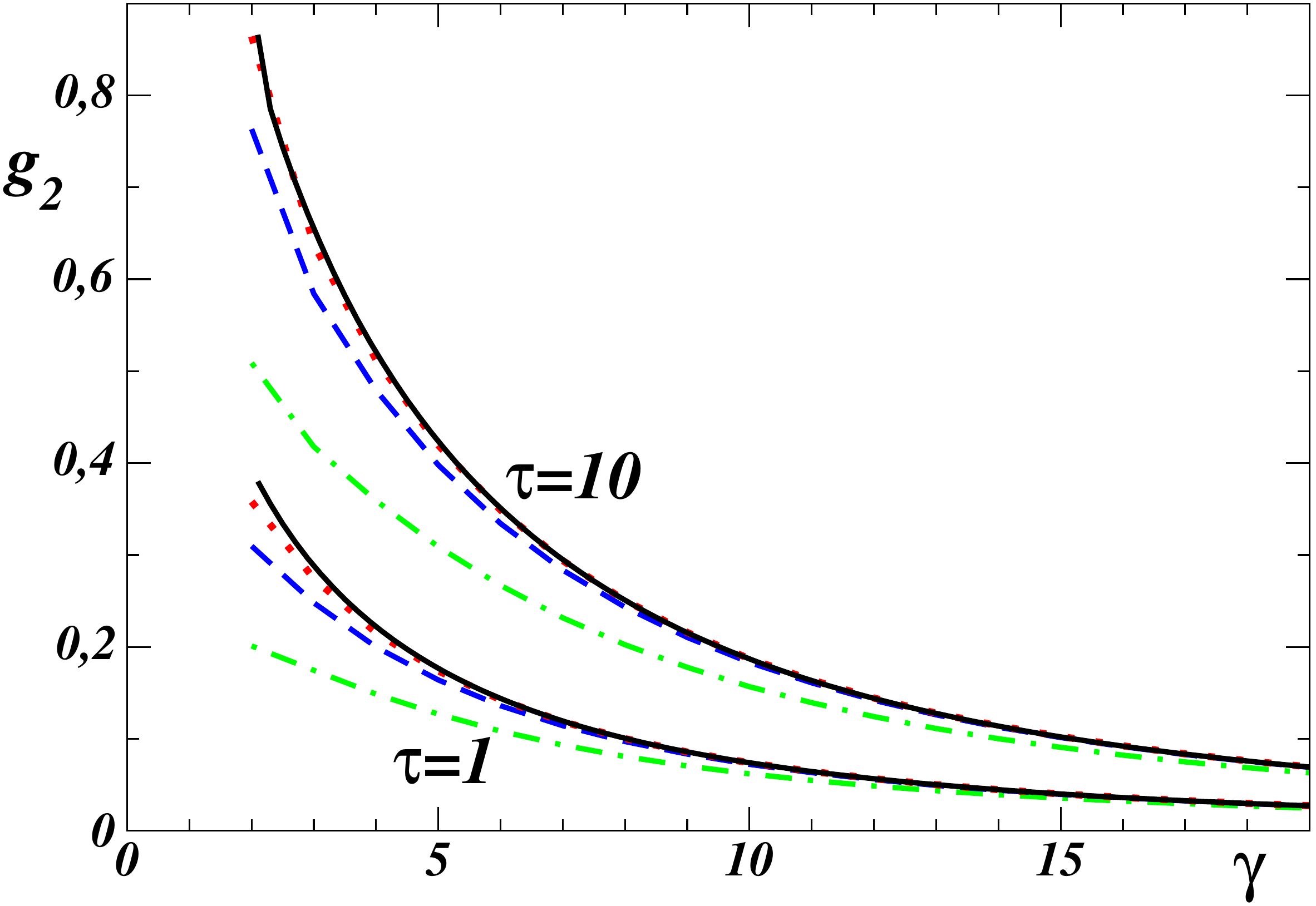}}
\caption{$g_2$ at $\tau=1$, $10$ using form factors up to $n=\textcolor{green}{4}$, 
$\textcolor{blue}{6}$ and $\textcolor{red}{8}$ particles with 
green dot-dashed, blue dashed and red dotted lines, respectively. The
solid lines show the exact result.}
\label{fig:fig2}
\end{figure}
As a second check of (\ref{eq:formula}) let us show how we can easily recover the leading order of the strong coupling (i.e.\ large $\gamma$) expansion of all $g_k$: since this always comes from the first non-zero integral in the series \erf{eq:formula}, we get 
\be
g_k=\frac{k!}{2^k}\left(\frac\pi\gamma\right)^{k(k-1)} I_k +\dots\,,
\label{eq:leading}
\ee
where 
$I_k=\int_{-1}^1\ud k_1\dots\int_{-1}^1\ud k_k \prod_{i<j}^k (k_i-k_j)^2$. This result coincides with the one obtained in \cite{gangardt}.

The quantity $g_2$ can be exactly determined via the Hellmann--Feynman theorem
\cite{kheruntsyan} and its plot at $T=0$ 
is shown in Fig.~\ref{fig:fig1} together with our determination from eqn \erf{eq:formula}. 
As before, also in this case there is a fast convergent behaviour of
the series. The strong coupling regime
of $g_2$ can be computed by expanding in powers of $\gamma^{-1}$ all the terms in eqn (\ref{eq:formula}) and for $T=0$ we get 
\be
g_2=\frac43\frac{\pi^2}{\gamma^2}\left(1-\frac6\gamma+(24-\frac85\pi^2)\frac1{\gamma^2}\right)+\mc O(\gamma^{-5})\,,
\label{strong-coupling}
\ee
in agreement with the result of the Hellmann--Feynman theorem
\cite{gangardt,kheruntsyan}. Expression \erf{strong-coupling} is also 
plotted in Fig.~\ref{fig:fig1} in order to show that 
the determination of $g_2$ (at finite $\gamma$)  obtained from the first terms of eqn \erf{eq:formula} is
closer to the exact result, because any of them contains
infinitely many powers of $\gamma$.  
At finite temperatures the convergence of the series is also pretty good and 
the results are shown in Fig.~\ref{fig:fig2}. 
\begin{figure}[t]
\scalebox{0.39}{\includegraphics{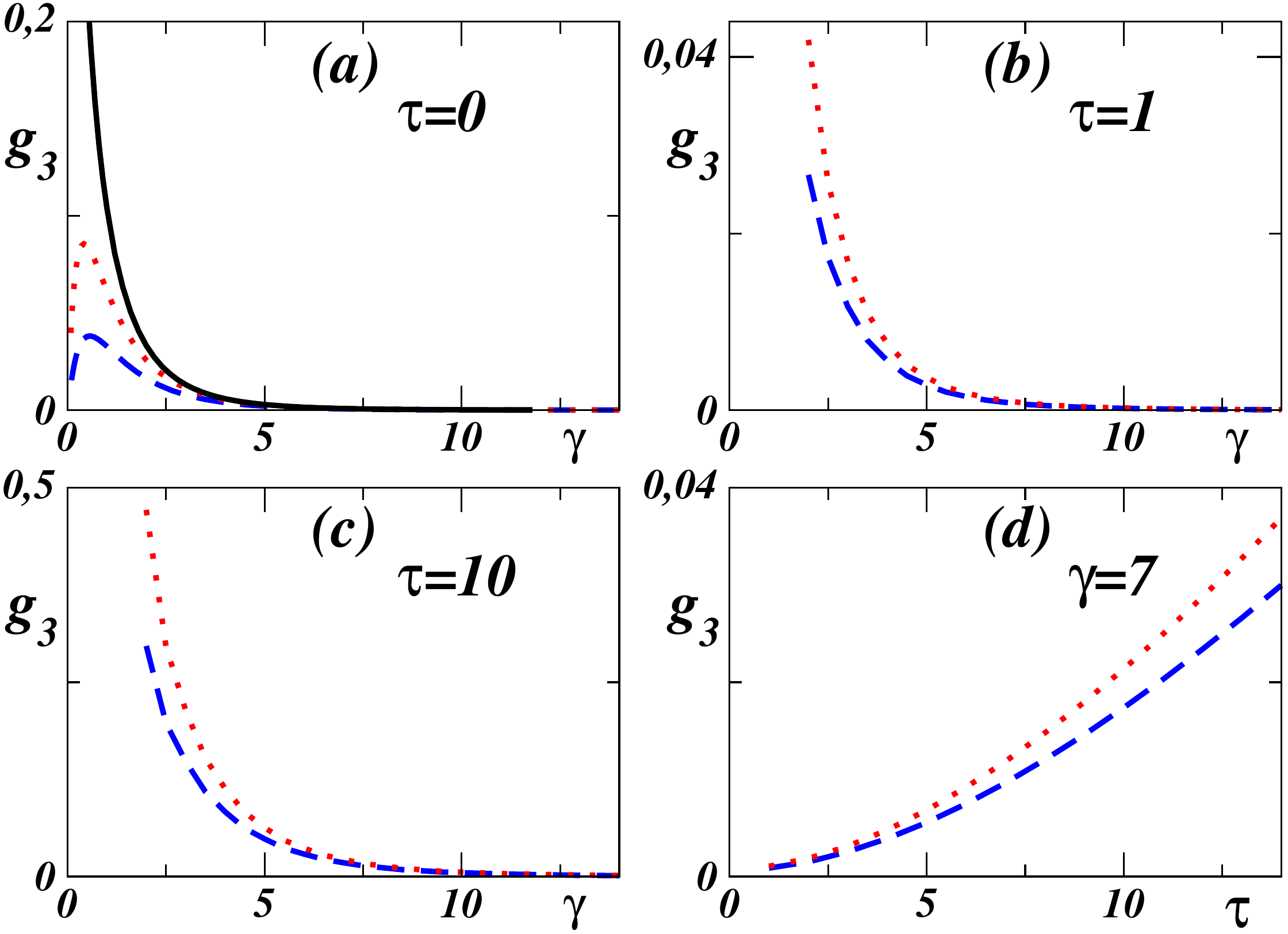}}
\caption{$g_3$ vs $\gamma$ at : (a) $\tau=0$,   (b) $\tau= 1$ and (c) $\tau=10$. In (d) 
we plot $g_3$ vs $\tau$ at $\gamma=7$. In all figures the blue dashed and the red 
dotted lines refer to $n=\textcolor{blue}{6}$ and 
$\textcolor{red}{8}$ particles, respectively. The solid line in (a) 
is the exact value of $g_3$ at $\tau=0$.}
\label{fig:fig3}
\end{figure}

As a final example, let us discuss $g_3$, 
a quantity known exactly at $T=0$ \cite{cheianov}, 
but only approximately at $T>0$ \cite{gangardt}. 
From (\ref{eq:formula}) its strong coupling limit at $T=0$  is 
\begin{equation}
g_3=\frac{16}{15} \frac{\pi^6}{\gamma^6} \left(1-\frac{16}{\gamma}
\right)+{\mathcal O} (\gamma^{-8})\,,
\end{equation}
where we report both the leading and sub-leading terms in
$\gamma^{-1}$ of this expression. The plot of $g_3$ at $\tau=0$ using
form factors up to $n=6$ and $8$ particles (i.e. one or two terms of
the series \erf{eq:formula}) is in Fig.~\ref{fig:fig3}(a) and, as in previous examples, it shows a nice convergent pattern to the exact value found in \cite{cheianov}. Figs.\ \ref{fig:fig3}(b,c) show $g_3$ as a function of $\gamma$ at fixed temperature $\tau$, while Fig.~\ref{fig:fig3}(d) shows instead $g_3$ as a function of $\tau$ at a fixed value of $\gamma$.

{\em Conclusion}. We have shown that the equilibrium expectation values for one-dimensional interacting Bose gases can be efficiently computed by using the non-relativistic limit of an integrable relativistic field theory, the sinh--Gordon model. There is a significant advantage in using this method instead of employing directly the non-relativistic Lagrangian (\ref{LagrangianNLS}). The reason is that a relativistic field theory 
presents a larger number of constraints (crossing invariance, for
instance) which permit to pin down exactly and efficiently the matrix elements of all operators: once these quantities are known, it is then easy to take their non-relativistic limit. As shown above, this proves to be a notable simplification in the computation of the correlators of the LL model. 

The method works equally well at $T=0$ and $T\neq 0$ where the series expansion presents a remarkable convergence behaviour for finite values of $\gamma$. There is no obstruction, in principle, to compute higher form factors and further improve the result. 
Strong coupling expansions in $\gamma^{-1}$ can be easily derived as well but 
at finite $\gamma$ the form factor expansion, containing
infinitely many powers of $\gamma$, is more accurate for the determination of $g_k$, as we showed 
comparing it with exact results.
As a significant application of the method 
we have determined $g_3$ at finite temperature (a term which is proportional to the recombination rate of the gas). This quantity, as well as the higher $g_k$, may  provide important information once the integrability  of the model is broken.
In the future it would be also interesting to apply this method both to 
two-point correlation functions and to other models.

\noindent
{\em Acknowledgements:} We would like to thank 
B. Pozsgay and G. Tak\'acs for discussions. 
This work is supported by the grants INSTANS (from ESF) and 2007JHLPEZ (from MIUR).

\end{document}